# Seismic structure of the crust in the western Dominican Republic

Diana Núñez a,b,*, Diego Córdoba b, Eduard Kissling c

a Centro de Sismología y Volcanología de Occidente (SisVOc), Universidad de Guadalajara, Mexico
b Dpto. Física de la Tierra y Astrofísica, Facultad de Ciencias Físicas Universidad Complutense de Madrid, Spain
c Institute of Geophysics, ETH Zürich, Zurich, Switzerland

A B S T R A C T

The contact between the Caribbean and North American plates is a tectonically complicated boundary where the deformation is accommodated in north and south of Hispaniola by the Enriquillo-Plantain Garden and Septentrional–Oriente Fault Zones (EPGFZ and SOFZ). We present a crustal and tectonic study of the Northeastern Caribbean Plate Boundary from wide-angle seismic data (WAS) acquired during the GEOPRICO-DO (2005) and CARIBE NORTE (2009) surveys, showing two transects crossing, from north to south, North American Plate (NOAM), Bahamas Platform, central Hispaniola and Caribbean Plate (CP). The results presented include two 2-D P wave velocity models of 425 km and 200 km long oriented NNE-SSW and ENE-SSW, respectively, obtained by the travel time forward and inverse modeling of the WAS data. Our study defines that the contact between Bahamas Platform, North American Plate and Hispaniola corresponds to oblique subduction with the Moho dipping 11° in the NNE-SSW direction. Furthermore, in the south, our results reveal the existence of an anomalous deep-reaching zone of lateral velocity variation in the mantle that could be associated with EPGFZ and a possible detached oceanic slab from NOAM that could explain the deep seismicity in the region.

## 1. Introduction

The region between the North American (NOAM) and the Caribbean (CP) plates is a 100–250 km wide seismogenic zone of mainly left-lat- eral strike-slip deformation that extends over 2000 km along the
northern edge of the Caribbean Sea (Mann et al., 1995) (Fig. 1a). The Puerto Rico Trench and the Virgin Islands are well-studied (Calais et al., 1992; Carbó et al., 2005; Bunce and Fahlquist, 1962; Dolan et al., 1991,
1998; Granja Bruña et al., 2009, 2010; Masson and Scanlon, 1991; among others) but the tectonic processes in the area of Hispaniola re- main poorly understood.

Different studies specifically investigated the tectonic setting of the region between the NOAM and the CP. Calais et al. (1992) suggested that the northern boundary of the CP is a transcurrent fault system, rather than a standard subduction zone. They describe it as a system that extends from Cuba to western Hispaniola and shows recent de- formation along the main left-lateral strike faults of Hispaniola and the highest topography of all Caribbean Islands (Calais et al., 2016). This model would explain the intermediate to deep seismicity of the zone
with the existence of a disconnected lithospheric slab and the trans- current faulting along boundary of the NOAM's oceanic lithosphere (Calais et al., 1992). Other studies postulate a complex dual subduction system whereby the NOAM oceanic slab dips southward along Hispa- niola Trench, with a strong vertical component down to 100 km depth due to the presence of the Caribbean slab along the Muertos Trough which dips northward 15° (see cross-section in Dolan et al., 1998). Better knowledge of crustal structure, Moho topography, and subMoho velocities could provide a more comprehensive understanding of the tectonic behavior in the eastern part of the NOAM-CP interaction zone in the Hispaniola Island region. This, in turn, would greatly enhance the overall tectonic understanding on which seismic hazard assessments are based in the region.

To address these issues, two multidisciplinary investigations were carried out in 2005 and 2009, GEOPRICO-DO and CARIBE NORTE projects, respectively. In this paper, we present the results of in-depth seismic investigations from the Bahamas Banks (Caicos, Mouchoir, Silver and Navidad Banks) crossing western Dominican Republic (DR) and reaching the northern part of the Beata Ridge.

## 2. Regional setting

Hispaniola comprises different oceanic plateau and island-arc ter- ranes with alternating mountain ranges and valleys, which are bounded by active left-lateral strike-slip, oblique-slip, or reverse faults (Fig. 1b) (Mann et al., 1991a). The northern part of Hispaniola is still colliding with the southern edge of the Bahamas Platform (Dolan et al., 1998; Calais et al., 2016), where the Bahamas Banks are located. This process takes place within the northern CP boundary region along a restraining bend (Mann et al., 1984) since late Miocene (Sykes et al., 1982) and is evidenced by compressional structures in the offshore region (Mann et al., 2002; Dolan et al., 1998). Due to its thicker-than-average crust, the Bahamas Platform resists subduction beneath the eastward-moving central and northern parts of Hispaniola (Mann et al., 1995). Large numbers of shallow earthquakes along the northeast margin of Hispa- niola indicate that this collision continues offshore to Navidad Bank and Silver Bank (Fig. 1b), indenting the northern Hispaniola margin (Bracey and Vogt, 1970).

Rising from 4 km depth, the Mouchoir Bank is relatively small ir- regular carbonate platform located over a regional basement. The Hispaniola basin is situated with a 3 km thick sedimentary cover between the northern Hispaniola slope and the Mouchoir Bank. Along the accretionary prism north of Hispaniola, an active fold and thrust belt was generated due to the oblique convergence of the Mouchoir bank along the Bahamas-Hispaniola collision zone (Mullins et al., 1992). In this area, the southeastern part of Bahamas is sliding ob- liquely beneath Hispaniola with a NE-SW maximum compression or- ientation attributed to tectonic tilting and subsidence. This motion may arise from the collision between NE Hispaniola and the banks located further southeast (Calais et al., 2016).

The northern part of Hispaniola is a crustal block constituted of an accretionary prism with recent material derived from Bahamas plat- form, serpentinite mélange of high-pressure metamorphic rocks that overlies an ophiolitic complex known as the Puerto Plata complex (PPC), and fragments of the intra-oceanic volcanic arc, and the North America continent (Escuder-Viruete et al., 2013, 2016; Hernáiz-Huerta et al., 2012). The Cordillera Septentrional is composed of units derived from the arc, the continental margin, and the intermediate oceanic li- thosphere, which were accreted during the convergence (de Zoeten and Mann, 1991). The Cordillera Septentrional is an elongated WNW-ESE mountain range of 15 to 40 km width, which reaches a maximum al-
titude of 1249 km and is bounded to the south by the Septentrional –
Oriente Fault Zone (SOFZ). The SOFZ extends onshore for 320 km across northern Hispaniola (Mann et al., 1998) and continues westward across the Windward Passage and the southern coast of Cuba (Calais et al., 1992). Eastward, this fault zone extends offshore and bifurcates into two major strands north of the Mona Passage (Dolan et al., 1998). The SOFZ is a left-lateral strike-slip fault responsible for faulting in the Cibao Valley area and for the uplift of the Cordillera Septentrional. This fault hosted a large historical earthquake 800 years ago, producing an oblique normal and left-lateral displacement of at least 4.6 m (Prentice et al., 2003). The offshore faults accommodate the convergent part of the transpressional NOAM-CP contact while the main onshore strike-slip faults accommodate the purely strike-slip component of deforma- tion. Both systems are separated by less than 100 km (Mann et al., 1998).

To the south, the Tireo formation and Peralta fold-and-thrust belt are the Cordillera Central's main geological structures (Hernaiz-Huerta and Perez-Estaún, 2002) and belong to the system known as Trans-

Haitian Belt. The first formation consists of volcanoclastic rocks be- longing to Circum-Caribbean Island Arc, a complex Late Cretaceous – Eocene system consisting of different tectonic units or terrains, which were initially formed and subsequently accreted within an intro-oceanic geodynamic environment since Late Jurassic – Lower Cretaceous (Lewis et al., 2002). The second one is a sequence of sedimentary rocks along the Cordillera Central southern flank in a NW-SE direction. The thrust of the basement over the Peralta fold-and-thrust belt is deformed by faulting and constitutes the contact between these units (Mann et al., 1991a).

Between the Sierra de Neiba and the Cordillera Central, the San Juan basin is filled by more than 7000 m of Tertiary and Quaternary sediments (Dolan et al., 1991; Mann and Lawrence, 1991). At its southeasternmost point, that basin connects with the Enriquillo basin, which together with the Azúa basin are part of the Hispaniola Neogene basins. The Azúa basin consists of a ramp basin filled with less than 3000 m of sediments whose geometry changed due to the north-eastward Pleistocene convergence of the Beata Ridge into south-central Hispaniola (Mann et al., 1991b). The Enriquillo Basin is a ∼5 km deep basin between blocks elevated by faults with oblique displacement. This system shows significant compression accompanied by little shear in an NNE-SSW direction and forms an elongate valley bounded to the south by active oblique (reverse and strike-slip) fault dipping to the south underneath the Sierra de Bahoruco (Symithe et al., 2015; Rodríguez et al., 2018). This valley was uplifted and altered by erosion creating some depressions such as Lake Enriquillo 42 m below sea level, which is gradually being lowered as a consequence of evaporation.

Recent studies indicate that the Enriquillo Plantain Garden Fault

Zone (EPGFZ) (Fig. 1a and b) extends from Jamaica to south-central Hispaniola and delimits the southern edge of the Gonave microplate (Calais et al., 2016; Symithe et al., 2015; Benford et al., 2012). The EPGFZ is forming a prominent and continuous lineament that extends from the Plantain Garden fault zone, through the Jamaica Passage crossing the southern peninsula of Haiti and reaching the Enriquillo Valley. This fault zone accommodates a main left-lateral motion and a small compressional component evidenced by the linearity of the fault and its association with folds, pull-apart basins, and bends (Koehler et al., 2013). This motion is accommodated on those folds and thrust belts distributed throughout the entire plate boundary rather than partitioned on the main bordering strike-slip faults (Corbeau et al., 2016a, b; Leroy et al., 2015).

1. Data acquisition and processing

Different studies have been carried out in the CP region (e.g., Officer et al., 1957; Ewing et al., 1960; Talwani et al., 1977; among others) but few of them have integrated onshore and offshore data to establish an internally consistent model for the reconstruction of the Caribbean tectonic history (Driscoll and Diebold, 1998). The GEOPRICO-DO and CARIBE NORTE projects (Carbó et al., 2005, 2010) encompassed mul- tidisciplinary data collected onshore and offshore of the Dominican Republic to achieve that goal. During both expeditions, the research vessel BIO Hespérides provided the marine seismic sources. In this work, we analyze the wide-angle seismic data recorded by land and marine equipment deployed along two profiles, A and F (Fig. 2). The offshore experiment did not record multichannel seismic data in the study area.

1.1. Experimental data from wide – angle seismic (WAS) project

The Profile A (Fig. 2) extends 425 km from Bahamas Bank (N) to Beata Ridge (S) crossing the western area of Dominican Republic (DR). The seismic sources used in this profile correspond to two marine shooting lines in the northern and southern extremes (LM1N and LM1S) and one borehole explosion (S3) of 1000 kg of explosives carried out during CARIBE NORTE fieldwork (2009). This profile was recorded by 97 TEXAN 125A (Trimble REF TEK, USA) land portable stations with one vertical-component geophone belonging to the IRIS-PASSCAL pool instrument (USA). These stations were installed from Pedernales to Puerto Plata (DR) with an inter-station interval of 2–3 km. Moreover, in this profile, three OBS were anchored in the Dominican Republic southern coast (Caribbean Sea), with the support of the patrol boat "Orion" provided by the Dominican Navy. The deployed OBS stations used in this study were LC2000SP short-period stations with three- component geophone sensors (L28), 4.5 Hz of natural frequency, and one hydrophone model HiTech HYI-90-U. OBS 11 was on shooting line while OBS 10 and 12 were 2 km to 2.5 km offline. A correction to the travel time, therefore, had to be applied before the interpretation of the results.

Eight portable seismic stations were deployed along the Profile F (Fig. 2) with an inter-station distance of 6 to 8 km. This seismic transect extends from Bahía de Neiba to the Azúa and Enriquillo Basins, crossing the western part of the Muertos Trough. The land stations recorded the L6 marine line shot during GEOPRICO-DO fieldwork (2005). The por- table seismic stations were composed of TEXAN 125A registering at 50 Hz combined with one vertical component geophone of 4.5 Hz, along with TAURUS (Nanometrics, Canada) and HATHOR3 (LEAS, France) digitizers, both registering at 100 Hz, combined with three- components seismometers LE-3D 1 Hz.

The airgun seismic source system used for both profiles, A and F, on board the BIO Hespérides consisted of two airgun subarrays with 7 Bolt® G-Guns 1500 L L and 1900 L L (3850 in3 of total volume). The shooting interval was 90 s, and the cruise speed was 5 knots. The seismic energy was recorded by the OBS with an offset of up to 100 km and by the land stations up to 200–300 km away. The processing steps for seismic data included merging with navigation data, corrections due to instrument drift and band-pass filtering, as well as the previously mentioned travel time corrections to the stations that were offset to the profile (Núñez et al., 2016).

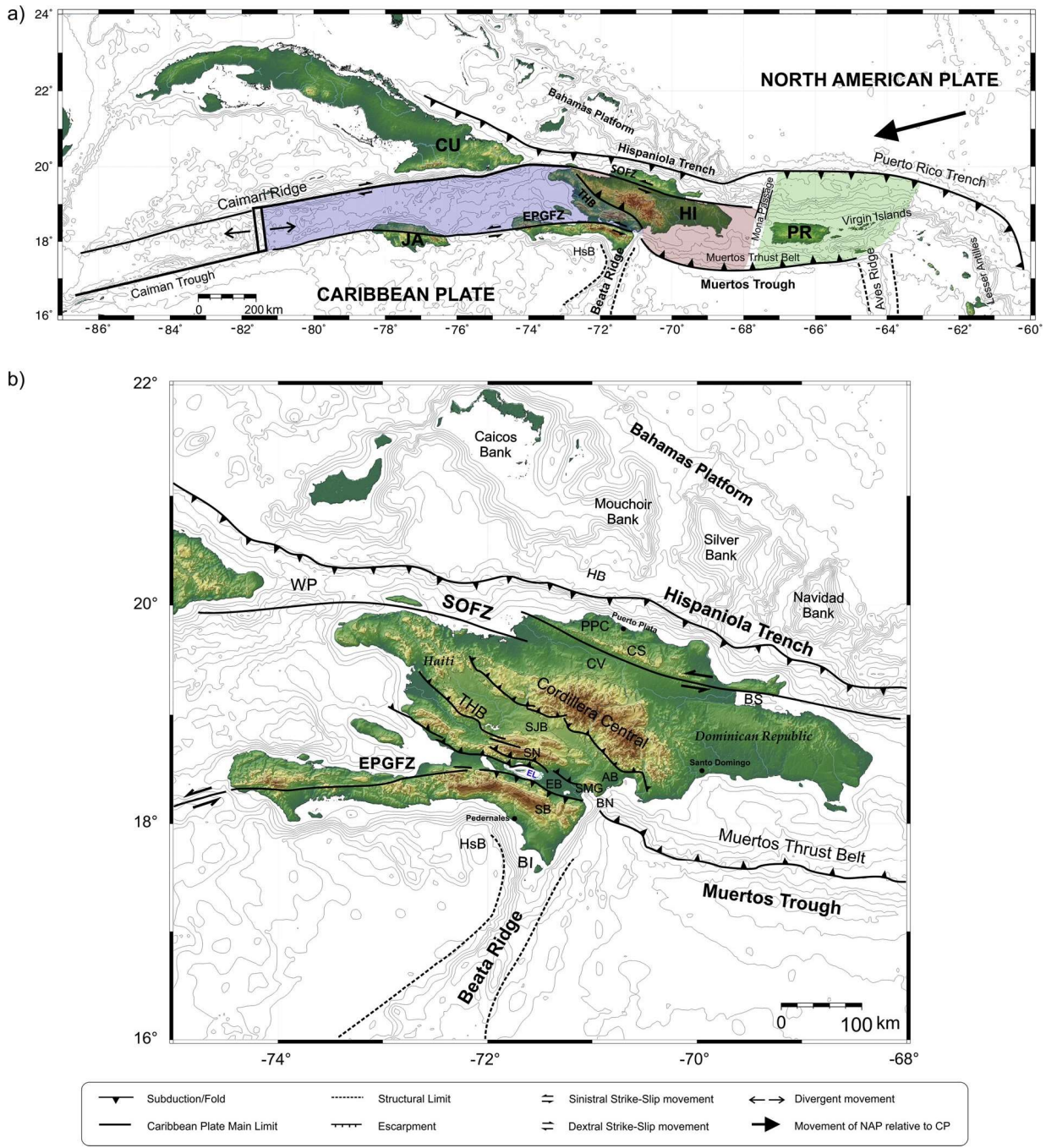

Fig. 1. (a) Tectonic frame of NE Caribbean border region (modified from Mann et al., 1999; Leroy et al., 2015; Pubellier et al., 2000; Corbeau et al., 2016a). Black arrow shows the relative displacement between the Caribbean and North American plates. From decreasing size, the Greater Antilles are Cuba (CU), Hispaniola (HI), Jamaica (JA), and Puerto Rico (PR). EPGFZ is the Enriquillo-Plantain Garden Fault Zone, SOFZ is the Septentrional - Oriente Fault Zone and THB is the Trans-Haitian Belt. Blue, red and green shadow zones represent the Gonave, Hispaniola and Puerto Rico – Virgin Islands microplates, respectively. (b) Schematic tectonic frame of Hispaniola with main tectonic and geologic features. AB Azúa Basin, BI Beata Island, BN Bahía de Neiba, BS Bahía de Samaná, CS Cordillera Septentrional, CV Cibao, EB Enriquillo Basin, HB Hispaniola Basin, HsB Haiti sub-basin, PPC Puerto Plata Complex, Valley, SB Sierra de Bahoruco, SJB San Juan Basin, SMG Sierra de Martín García, SN Sierra de Neiba, WP Windward Passage. (For interpretation of the references to colour in this figure legend, the reader is referred to the web version of this article.)

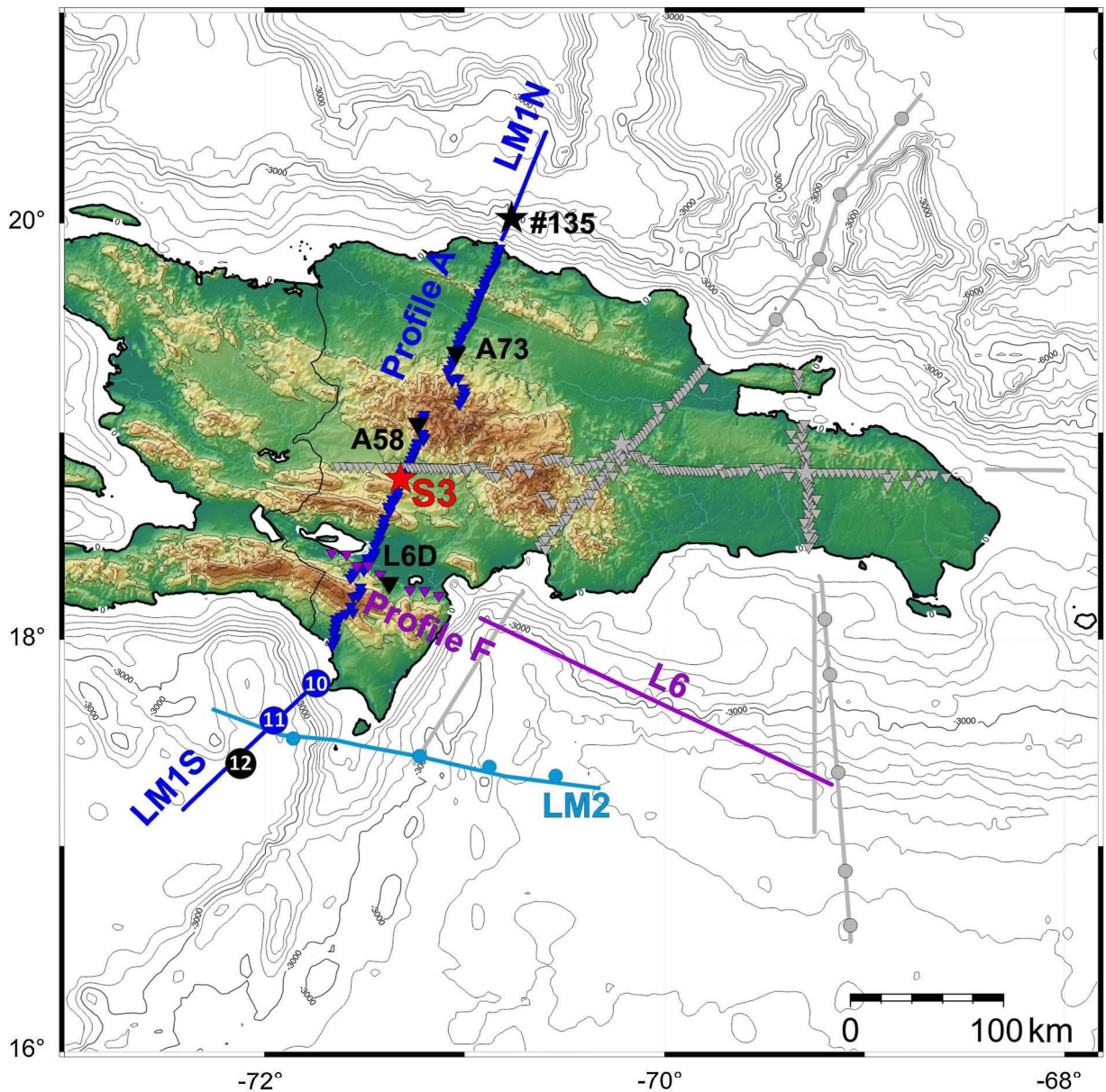

Fig. 2. Location map of stations and seismic sources used in this paper (colored lines and symbols) and all stations deployed during the CARIBE NORTE and GEOPRICO-DO projects (grey symbols) (Núñez, 2014). Black symbols represent the seismic station and sources locations whose record sections are shown in Section 2.1. Light blue line depicts the location of Profile LM2 (Núñez et al., 2016). (For interpretation of the references to colour in this figure legend, the reader is referred to the web version of this article.)

During GEOPRICO-DO and CARIBE NORTE cruises, bathymetry data were acquired with Kongsberg Simrad EM 120 multibeam echosystem and included throughout the interpretation and modeling steps of the seismic data.

### 3.2. P-wave phases picked

In this work, we present some selected record sections obtained for both profiles. In these selected sections, it was possible to identify time

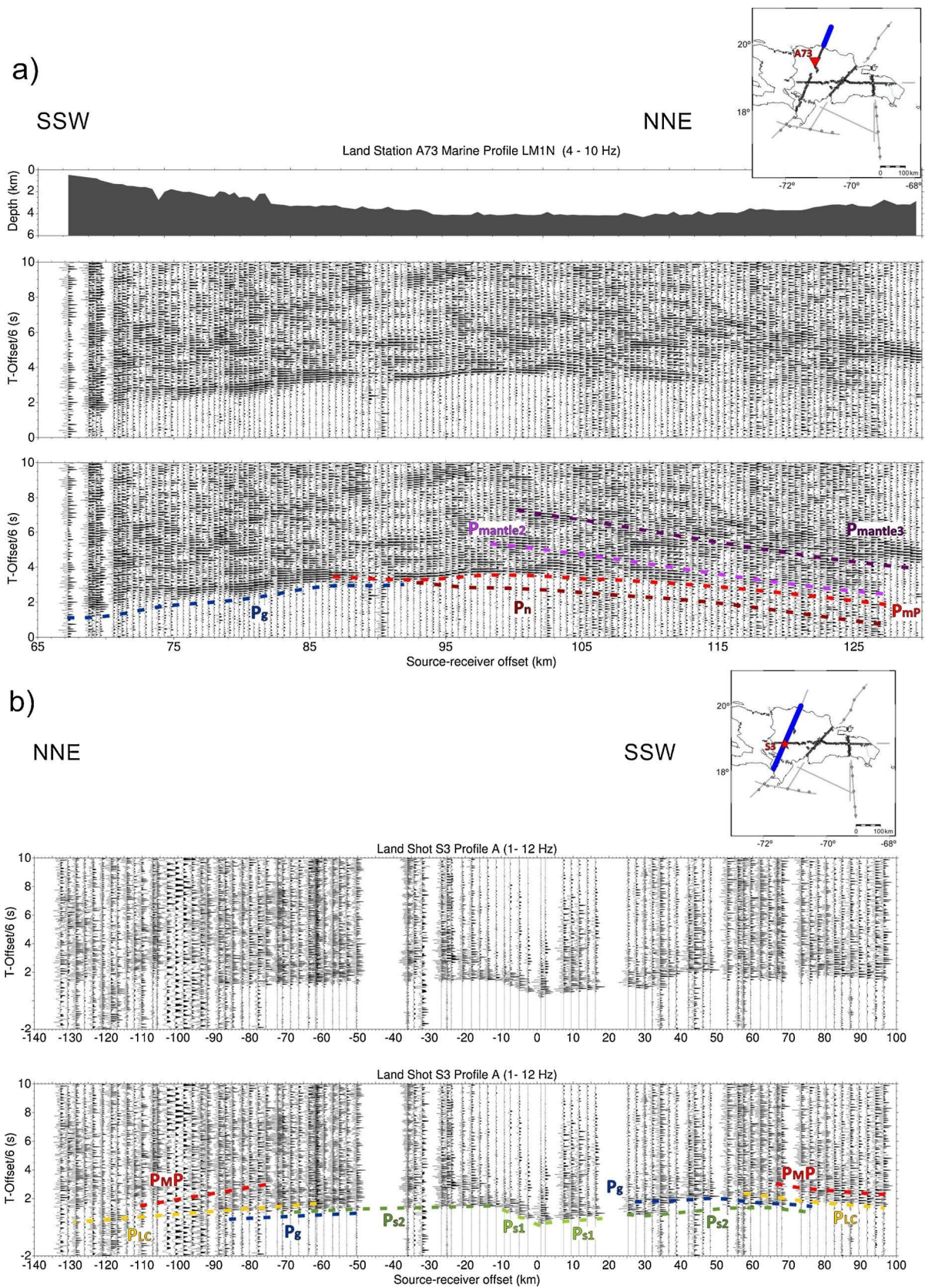

Fig. 3. Record sections (vertical component) of: (a) A73 in seismic line LM1N, (b) Land shot S3 recorded by seismic stations of Profile A, (c) Single shot 135 of seismic line LM1N recorded by seismic stations of Profile A, (d) A58 station in seismic line LM1S, (e) OBS12 in seismic line LM1S, and (f) L6D station in seismic line L6 (GEOPRICO-DO Project). The reflected and refracted P-wave horizons are depicted by color dashed lines. On top of every seismic record section is shown the bathymetry along the marine transect. Band-pass filtered from 4 to 10 Hz applied in a, d, e and f and from 1 to 12 Hz in b and c. Amplitudes are trace normalized and the time axis is reduced by 6 km/s. Labels of the P-wave phase correlations are $P_w$, $P_{S1}$, $P_{S2}$, $P_g$, $P_{LC}$, $P_MP$, $P_n$, $P_{mantle1}$, $P_{mantle2}$ and $P_{mantle3}$. The vertical scale is common to all record sections, but the horizontal scale is adjusted for each section. (For interpretation of the references to colour in this figure legend, the reader is referred to the web version of this article.)

c)

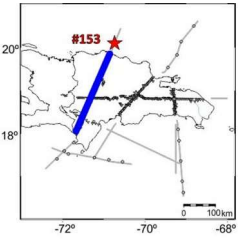

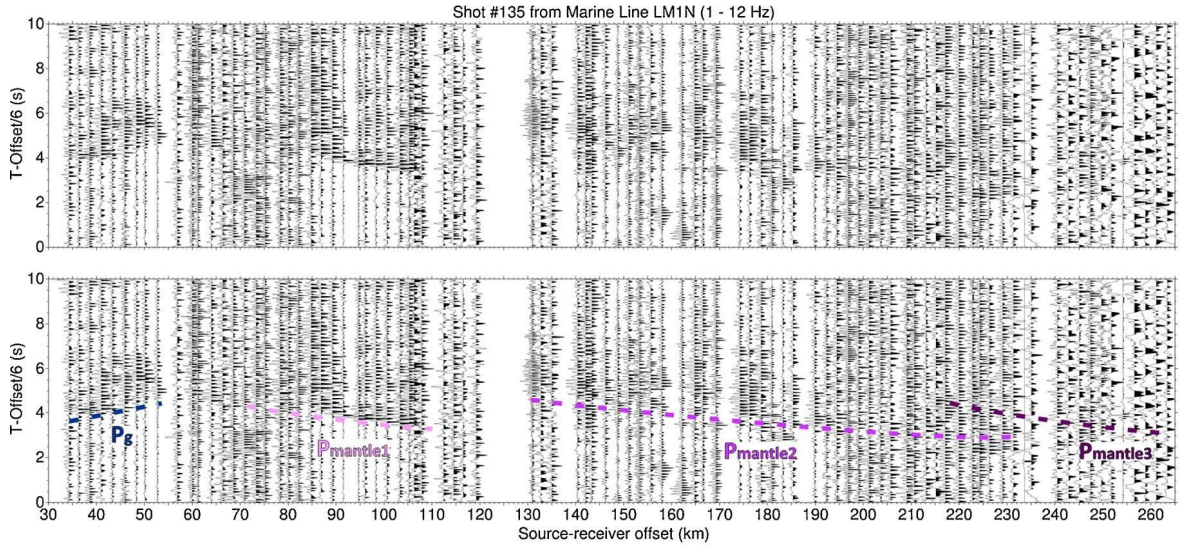

d)

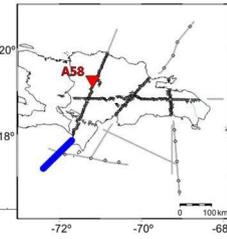

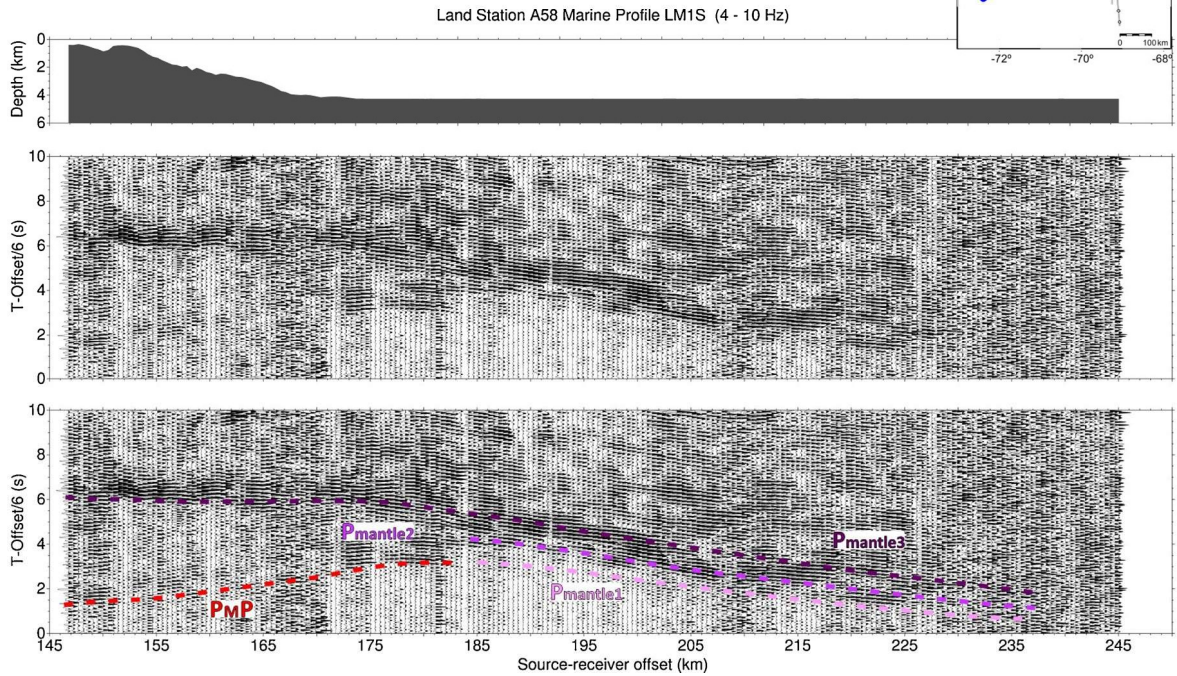

Fig. 3. (*continued*)

correlations corresponding to P-wave refracted and reflected phases in the crustal and upper mantle discontinuities and the next multiple of these seismic phases. We have interpreted $P_{S1}$ and $P_{S2}$ phases traveling through the sedimentary cover in the upper crust. The $P_g$ is a refracted wave going in the middle crust (presumably composed of mafic rocks) and in the lower crust (likely composed of gabbros). The $P_{LC}$ is a reflected phase at the top of the lower crust. The $P_MP$ is the reflected phase at the Moho discontinuity; $P_n$ is a refracted phase on the upper mantle; $P_{mantle\#}$ (# = 1, 2, 3) are phases that are reflected in upper mantle discontinuities.

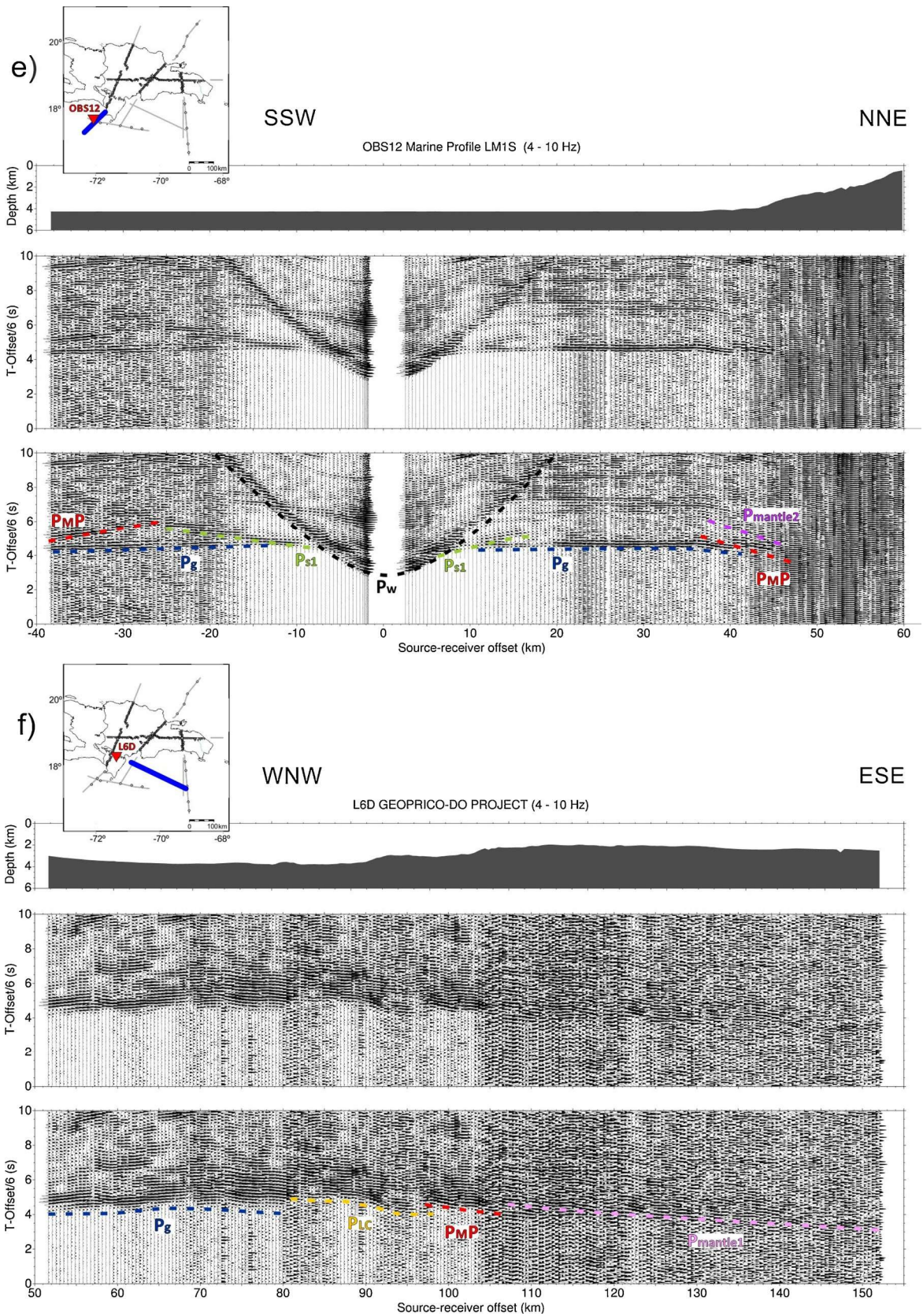

Fig. 3. (*continued*)

As an example of the work carried out, we selected the most representative six seismic record sections to illuminate the subsurface structure and their relation to tectonic features crossed by these profiles. The first record section corresponds to A73 land station recording the marine seismic line LM1N (Fig. 2). This station is located 130 km away from the farthest shot of the seismic line in the region of the Cibao Valley (Fig. 1b). Fig. 3a depicts the P-wave phases interpreted in this record section. $P_g$ phase is observed from 68 to 93 km of source-receiver offset with 4.5 km/s of average apparent velocity (*aav*). The substantial impact of bathymetry in the area can be identified in the shape of the observed $P_g$ phase. Between 91 to 127 km offset, $P_MP$ and $P_n$ are correlated with 10.8 km/s of *aav* for $P_n$ phase, indicating that the Moho is dipping to the south. The $P_{mantle2}$ and $P_{mantle3}$ phases are reflected phases in the upper mantle, which can be correlated from 107 to 125 km and 105 to 130 km offset range, respectively.

Hispaniola is seismically sampled by the S3 land shot and some selected marine shots, which were registered by all seismic stations deployed along Profile A. Fig. 3b, and c shows two examples of the obtained sections. In Fig. 3b, we can observe the seismic record section of S3 borehole explosion, located in the San Juan Basin (Figs. 1b and 2) where it is possible to correlate phases traveling through the first sedimentary layer down to 12 km depth and down to 16 km in the northern and southern parts of the transect with *aav* of 3.8 and 5.5 km/s, respectively. These phases are followed by refracted phases in the second sedimentary layer up to 70 km offset in the region of Cordillera Central and 75 km from the shot in the Sierra de Bahoruco (Fig. 1b). The $P_g$ phase was identified between 50–90 km of offset to the NNE and 30–70 km to SSW of the seismic record section, respectively, with 6.4 km/s of *aav*. In the NNE profile segment $P_{LC}$, a reflected phase in a layer of the lower crust, is identified in an offset range between 60 and 130 km. In this section of the profile, the $P_MP$ phase is correlated from 75 to 113 km offset. Between 90 to 105 km source-receiver offset, it is possible to observe a decrease in the signal/noise ratio that could be associated with the presence of the SOFZ. In the SSW part, the $P_{LC}$ is observed between 55–97 km offset and the $P_MP$ is identified between 77–97 km offset.

The marine shot 135 is located 60 km from the first shot of the LM1N line (Fig. 2), and its seismic record section (Fig. 3c) illustrates the deep structure across Profile A onshore transect. In this seismic record section, different reflected phases in the upper mantle, $P_{mantle1}$, $P_{mantle2}$, and $P_{mantle3}$, are correlated in the offset ranges 75–112, 130–235 and 225–265 km, respectively.

The southern region is sampled by different land stations and three OBS registering the LM1S marine line (Fig. 2). As an example, Fig. 3d shows the record section corresponding to A58 station deployed in the Cordillera Central area (Fig. 1b) at 192 km from the farthest shot of LM1N seismic line. In this section, we interpreted four deep reflection phases. The first one is $P_MP$ observed from 146 to 176 km offset while $P_{mantle1}$ and $P_{mantle2}$ are interpreted in an interval of 176 and 230 km source-receiver offset. The most in-depth phase is $P_{mantle3}$ registered in almost the whole record section (146–230 km).

At 65 km to the southern coast of the Dominican Republic in the Caribbean Sea region, the OBS 12 was anchored in the Haiti Sub-basin at 4.3 km below sea level (Fig. 2). Fig. 3e displays the P-wave phases of this seismic record section. The first phase identified in this record section was interpreted as a $P_{S1}$ from 6 to 25 km offset towards the SSW and to 17 km toward the NNE directions of the seismic record section, displaying an *aav* of 5.0 and 4.7 km/s, respectively. Then, the $P_g$ is interpreted between 11 and 42 km offset range with 6.0 km/s *aav* to the NNE and from 13 to 40 km offset with 6.2 km/s *aav* to the NNE. From the OBS to the Haiti Sub-basin, the most in-depth phase observed is a $P_MP$ from 25 to 40 km, and towards to coast, it is interpreted $P_MP$ and $P_{mantle1}$ phases in offset ranges of 38–48 and 39–46 km, respectively.

The portable seismic station L6D is located between Sierra de Bahoruco and the Enriquillo Basin (Fig. 1b) and registered the marine seismic line L6 with WNW-ESE direction shot during GEOPRICO-DO experiment (Fig. 2). First arrivals correspond to a $P_g$ between 50 and 80 of source-receiver offset with 5.6 km/s of *aav*. From 80 km offset, it is observed the increasing noise level that could be due to the presence of Muertos Thrust Belt and where we can interpret $P_{LC}$ phase up to 95 km, a $P_MP$ from 97 km to 106 km and one deeper phase, $P_{mantle1}$, from 97 to 150 offset range.

### 3.3. Uncertainty estimation of phase picking and P-wave velocity model

The final P-wave velocity model uncertainty is obtained from an evaluation of different uncertainties that affect the modelling process (Núñez et al., 2011). These uncertainties have been calculated applying the procedure suggested by Núñez et al. (2016) and are mainly due to i) Picking of arrival times, and ii) Differences between seismometer position and seismic line offset, that can be applied to iii) P-wave velocity model errors in velocity and depth.

In our case, we have two types of seismic record sections: i) a single seismic station recording one marine seismic transect; and, ii) one shot recorded by a line of seismic stations. In most of them, those errors were estimated for each phase as a mean value of the single estimation of errors due to the high number of traces interpreted as the same seismic phase. We obtained a maximum value for the arrival time picking uncertainty of 0.09 s.

The calculations made by Núñez et al. (2016) about the relative position between the seismic transect and the instruments show that the error is within the uncertainty range of phase picking for offset distances less than 2 km. Most of the stations used for modeling in this study are in this assumption. For the stations with higher offsets, the error estimation was added to phase picking uncertainty, obtaining a maximum value of 0.13 s.

In order to obtain the depth and velocity error estimation in a velocity model, we also used the methodology of Núñez et al. (2016). In this study, we respectively obtained a range of 0.2 to 0.8 km and 0.1 km/s as depth and velocity uncertainties for the shallowest part. For the deepest regions, these quantities increase up to 3 km and 0.3 km/s, respectively.

### 3. Controlled-source seismology (CSS) modelling results

The interpretation of crustal wide-angle reflection and refraction data to derive a velocity structure is a laborious trial-and-error forward modeling process that involves 2D ray tracing. The principles applied in this study are based on the repeated comparisons between theoretical travel time and amplitude response of a laterally inhomogeneous medium with observed seismic record sections. The process is repeated until an adequate model matching observed and calculated parameters is obtained (Zelt and Ellis, 1988). In this study, the final 2D velocity and interface structure were obtained by forward-modeling and subsequent travel time inversion using the software package (TRAMP, RAYINVR, and DMPLST) developed by Zelt and Smith (1992).

According to Zelt and Smith (1992), this method can be applied to any travel time data set, regardless of the quality of seismic data and the geometry between the seismic source and the receiver, whenever forward modeling is possible. For this method, each layer has the position and number of velocity and boundary depth nodes specified, which should be suitable for maximum ray coverage. This model parameterization allows the inverse technique to be more effective. The initial forward modeling step uses a robust ray tracing method with the simulation of smooth layer boundaries to increase the stability in the inversion. Finally, to constrain the velocity contrast for reflected waves across the layer boundaries and the velocity gradients in the layers for refracted waves, synthetic seismograms were calculated using the programs included in the Zelt's software package (PLTSYN) (Zelt and Smith, 1992).

Phase correlation was performed for a selection of the most informative land, and marine shot seismic record sections of profiles A and F across the Western Dominican Republic (Fig. 2). We used 40 seismic records for profile A and 5 for profile F, for which interpretation was based on determining the observed P wave phases, as well as their apparent velocities. Both provided the necessary information to generate an initial velocity distribution and depths of seismic limits in the crust and uppermost mantle. Afterward, models of ray tracing and synthetic seismograms were calculated repeatedly to finally obtain a suitable model that fits the observed data within the error limits. In this study, we used the $\chi^2_N$ parameter to check the goodness of fit between observed and theoretical data. According to Afilhado et al. (2008), an ideal velocity model should provide a $\chi^2_N$ value equal to 1, while if the data is over-fit $\chi^2_N$ is less than 1, otherwise, quantities more than 1 are obtained for under-fit data. In practice, the $\chi^2_N$ values are substantially different from 1 (Zelt and Forsyth, 1994; Zelt, 1999). The proposed models were also evaluate using another two quality parameters: i) the percentage of data points that are hit by rays traced in the model, and ii) the RMS travel time error.

For our models, we took elevation and water-depth values from a DEM of the Hispaniola Island and bathymetry and navigation data provided by the BIO Hespérides, respectively. Regarding the model distance of the investigated transects, we selected a point located 69 km north of the land station A101 for profile A and 35 km west of the land station L6D for profile F (Fig. 2).

### 3.1. Profile A

The final velocity model of Profile A (Fig. 4a) has 6 layers and 3 reflectors, corresponding to sediments, middle and lower crust, and upper mantle, and they were constrained by the data mostly by reflections from first order velocity discontinuities. It is restricted by travel time and controlled by amplitudes using synthetic seismograms. The relative amplitudes study exposes that observed and predicted data are similar within the range of uncertainty (See supplementary material S2). This model reproduced 93% of observed travel times. For each phase interpreted, we estimated the arrival-time fit quality ($\chi^2_N$) in profile A with most of individual phases under-fitting ($\chi^2_N > 1$). Our final model for Profile A produces a $\chi^2_N$ of 1.8, that is close to the ideal case (Table 1).

Profile A is 425 km in length whose northern part is a regular oceanic crust with a thickness of 7–8 km below a water layer of 3.1–4.1 km thick with a P-wave velocity of 1.5 km/s and thin sedimentary layers (< 1 km of thickness) with a $V_P$ increasing from 3.3 to 5.4 km/s, with ∼ 0.1 km/s of uncertainty. The basaltic layer bottom is situated at 5 km depth while the Moho discontinuity is reached at 10 ± 1 km and it increases with model distance (See supplementary material S1).

At 90 km model distance, the northern coast of Hispaniola is reached and sedimentary layers increase in thickness up to 3 km. Moreover, the basement is 4 km thick, with depth uncertainty less than 1 km. This zone corresponds to the Puerto Plata Complex (Fig. 1b). It is followed by the western flank of Cordillera Septentrional whose effect is not very pronounced in the seismic record sections. Sediments beneath the Cibao Valley (Fig. 1b) are 4 km thick, and along its southern flank, the granitic basement rises to the surface. In this area, the Moho is 27 ± 2 km deep.

In the San Juan Basin, the sedimentary cover reaches a maximum depth of 5 km. In the Sierra de Neiba, basement velocities are only found at greater depths than in other mountain ranges, providing good evidence for the presence of volcano-sedimentary rocks at shallow levels. In this area, Moho is poorly constrained but could reach the greatest depths along the model with values close to 32 km.

The area of Enriquillo Basin that Profile A crosses (Figs. 1b and 2) is situated below sea level and filled by alluvium deposits. The seismic stations located in this region show a low S/N ratio in their seismic signals. Despite this, seismic phases could be correlated whose modeling allowed obtaining a maximum value of 5 km for the basin depth. Further south, the Sierra de Bahoruco's basement rises again reaching almost subsurface, but in the southern coast, it deepens up to 7 km with velocities of 5.8–6.0 ± 0.2 km/s.

The deep structure including lower crust, Moho boundary, and upper mantle shown in this profile (Fig. 4) indicates clear differences between north and south. In the northern part, the interface between NOAM crust and the basement of Hispaniola Island dips 11° towards to the south with a P-wave velocity contrast of 6.6 ± 0.1 km/s and 7.6 ± 0.2 km/s in Moho discontinuity. The velocity in the upper mantle increases slowly between 7.6 and 8.0 km/s within a layer of an average thickness of 20 km.

Offshore in the Caribbean Sea, the water depth practically remains constant at 4.7 km, and the first sedimentary layer is approximately 2.5 km thick. Below OBS 11, the Moho rises to 19 ± 2 km depth with an increasing uppermost mantle velocity with respect to northern part of profile, constituted by the original oceanic crust and the thick oceanic plateau (Mauffret and Leroy, 1997). The $V_P$ contrast in the next layer below the Moho boundary is 7.8 and 8.3 km/s, obtained from numerous reflected rays (Fig. 4b). Towards the SSW, the model presents an oceanic crust with a sedimentary cover of 4 km and 10 km of the total thickness. The deepest floating reflector interpreted is found from 40 to 155 km model distance with a depth between 45 and 75 ± 3 in the north, and from 275 to 340 km model distance reaching up to 72 ± 3 km in the south with a contrast of $V_P$ between 8.3 and 8.5 km/s ( ± 0.2 km/s) (Fig. 4b).

Between the northern and southern parts of Hispaniola along EPGFZ, an anomalous zone with lateral variation in velocity (from 7.5 ± 0.2 km/s to 7.0 ± 0.2 km/s) is observed though poorly resolved (Fig. 4a, supplementary material S2). In this area, a more complex structure is found and associated with a variation in the seismic energy. This structure causes dispersion and attenuation of energy, especially at the A58 station record section (Fig. 3d).

Additionally, we have verified the accuracy in determining the Moho boundary depth from different lower crust velocities in the northern region (6.3, 6.7, and 7.0 km/s) and in the southern region (6.4, 6.8, and 7.0 km/s). The results for the northern region (Table 2) indicate that the onshore region where PPC, CS, and CV are located have higher differences than the offshore and the Cordillera Central regions. However, in the southern region, it is observed that the zone of the sub-basin of Haiti has fewer differences than the southern onshore region. The maximum values obtained concerning the model proposed in this work correspond to the lower crust velocity of 7.0 km/s.

### 3.2. Profile F

Profile F (Fig. 2) is 300 km long, and it was shot along a WNW – ESE direction. The record sections generally exhibit a high signal-noise ratio (Fig. 3f) revealing the significant subsurface features down to 45 km depth, because of that, it was not possible to evaluate the uncertainties of this model. The final velocity model of this profile (Fig. 5a) has five layers and one floating reflector constrained by the data (Fig. 5b). This model fits 97% of observed travel times within their respective uncertainty estimates (Table 1).

On land, the shallow structure is poorly resolved, but it is possible to infer two thin sedimentary layers with an increasing P-wave velocity (3.3–4.8 km/s), based on data of the previous section, where the basement is located. This sedimentary layer could be 1 km thick each disappearing between 50–80 km from the origin of the model and being slightly thicker offshore. The bottom of the upper crust is located onshore and offshore at about 5 km depth. Although we have only non-reversed record sections, we could establish that the thickness of the upper crust and sedimentary layers are significantly reduced in the offshore region with an increase in the velocity when compared to the model of Profile A. Middle, and lower crust are delineated as being thicker than upper with P-wave velocities of 5.8 and 6.6 km/s. The middle crust rises to subsurface (115–145 km) due to the Muertos Trough presence (Fig. 1b).

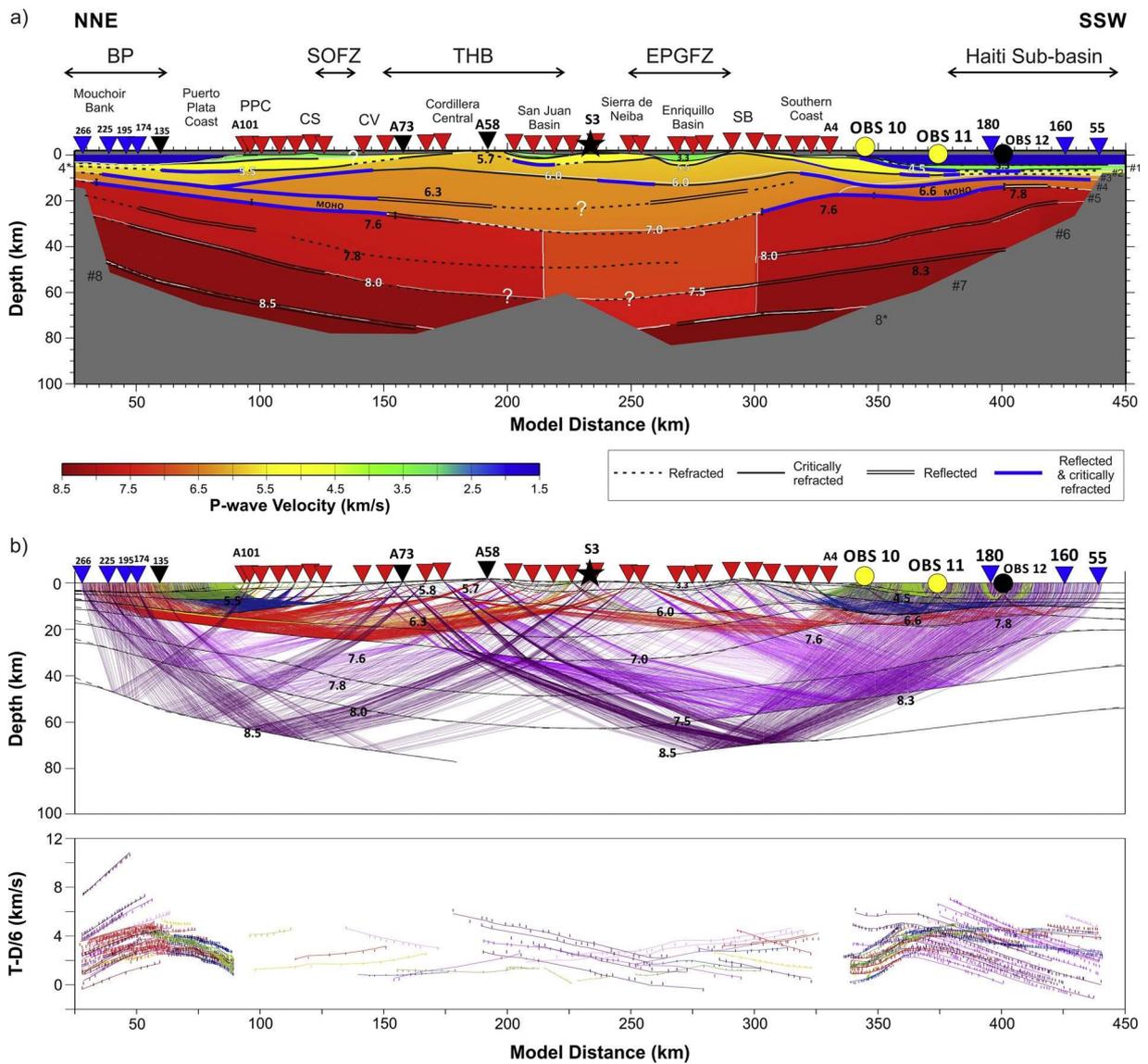

Fig. 4. (a) P-wave velocity model across the western part of the Dominican Republic (CARIBE NORTE Project). Seismic stations of Profile A are depicted by red inverted triangles and yellow circles. Black symbols represent the seismic station and sources locations which record sections are shown in Section 3.2. Blue inverted triangles represent single shots extracted from marine seismic profiles LM1N and LM1S. Black star represents land shot S3. Horizontal and vertical axis show model position and depth below 0 in km, respectively. The colored area shows the domain where ray tracing limits velocities. The lines represent the boundary of the layers, while the positions where the rays are refracted, critically refracted, reflected and both simultaneously are marked by pointed, solid, double and blue thick lines, respectively, showing the well-defined areas. The uncertainty in depth of Moho discontinuity is represented by vertical bars. The region not crossed by rays is represented by a grey area, while the question marks show regions with poor control over the ray coverage. The layer numbers are identified by #N, where N is the layer number, and N* depicts reflectors. BP Bahamas Platform, CS Cordillera Septentrional, CV Cibao Valley, EPGFZ Enriquillo – Plantain Garden Fault Zone, PPC Puerto Plata Complex, SB Sierra de Bahoruco, SOFZ Septentrional – Oriente Fault Zone, THB Trans-Haitian Belt. (b) Top: Ray tracing and velocity model with average velocity in the corresponding layer in km/s. Bottom: Comparison between observed (vertical bars) and calculated (lines) travel times. Its height is the uncertainty estimated for every phase. In all plots, distances refer to the velocity model origin. (For interpretation of the references to colour in this figure legend, the reader is referred to the web version of this article.)

The analysis of PMP phase reveals that the Moho discontinuity is 27 km deep (at 70 km offset in the model) increasing up to 32 km (at 90 km in the model) and immediately rises to only 17 km at the Muertos Trough. The velocity changes from 6.6 to 7.6 km/s at this discontinuity. The most in-depth phase found in the data corresponds to a floating reflector with a velocity contrast of 7.6–8.1 km/s located at 43–47 km depth (Fig. 5b).

The quality estimation of arrival-time fit ($\chi^2_N$) for every interpreted phase in this study are shown in Table 1. The final Profile F model originates a $\chi^2_N$ of 1.2, not far from the perfect case that could indicate that observed arrival-times obtained during GEOPRICO-DO are not enough to constrain a precise velocity model.

4. Discussion

The analysis of refraction and wide-angle reflection seismic data obtained in GEOPRICO-DO, and CARIBE NORTE projects, reveals tectonic differences between north and south in the crustal structure of the western Dominican Republic. The upper and middle crustal structures vary significantly throughout the profile. In the northern region, our study does not reveal significant changes in the crust and upper mantle associated with the SOFZ in the vicinity of the western flank of the Cordillera Septentrional. This lack of information can be attributed to the perpendicularity of our profile relative to this structure and the restrictions of the WAS method, which could be overcome by combining it with near-vertical reflection or P-wave delay techniques.

Table 1
Travel time fit. Number of picks (NC), Pick uncertainty (Sigma(s)), Number of traced rays (NR), Travel time root mean square misfit (Trms), and normalized Chi-square ($\chi_N^2$) for each phase.

| | Phases | NC | Sigma (s) | NR | Trms (s) | $\chi_N^2$ |
|---|---|---|---|---|---|---|
| PROFILE A | Ps1 | 269 | 0.095 | 256 | 0.126 | 2.224 |
| | Ps2 | 342 | 0.102 | 328 | 0.137 | 2.057 |
| | Pg | 693 | 0.115 | 672 | 0.135 | 1.648 |
| | PLC | 74 | 0.124 | 70 | 0.081 | 0.616 |
| | PMP | 610 | 0.128 | 573 | 0.141 | 1.551 |
| | Pn | 263 | 0.099 | 214 | 0.150 | 1.712 |
| | Pmantle1 | 274 | 0.123 | 228 | 0.179 | 2.560 |
| | Pmantle2 | 478 | 0.115 | 458 | 0.161 | 1.954 |
| | Pmantle3 | 402 | 0.132 | 374 | 0.162 | 1.890 |
| | FIT | 3405 | 0.116 | 3173 | 0.148 | 1.824 |
| PROFILE F | Pg | 59 | 0.143 | 58 | 0.144 | 0.924 |
| | PLC | 41 | 0.151 | 41 | 0.182 | 1.000 |
| | PMP | 28 | 0.146 | 24 | 0.176 | 1.152 |
| | Pn | 11 | 0.155 | 11 | 0.161 | 1.185 |
| | Pmantle1 | 36 | 0.162 | 36 | 0.259 | 2.119 |
| | FIT | 175 | 0.150 | 170 | 0.188 | 1.228 |

The Moho deepens through the island interior from north to the south showing a dip angle of 11°. Our data indicate that the crust could extend to 27 km depth in the region of the Cordillera Central. Since the resolution of the method is not able to distinguish between the transitional and the oceanic NOAM Moho, the crust may be associated with the transitional crust of the island. Besides, as the Cordillera Central is the mountain range with the highest elevations of the Antilles reaching 3100 m at the Pico Duarte, this thickening could be associated to the lithospheric load provided by this mountain system. From the densities (2.8, 3.2, 3.4 g/cm$^3$) obtained from the P-wave velocities shown in our model (6.3, 7.6–7.8, 8.3 km/s) using the Nafe and Drake equation (Ludwig et al., 1970), it is possible to estimate the root depth of the said mountain range of ∼ 15 km. Assuming a 20 km thick transitional crust in this region, the total crustal thickness would be ∼ 38–40 km. Once, compared to the 27 km proposed in this study, it can be concluded the tectonic processes related to the convergence between the NOAM and the CP prevent the isostatic equilibrium of the system.

From south to north, the contact between the CP and the southern Hispaniola presents an average dip angle of 8°-9°. In the southern part, the Moho depth decreases to 13 km, where the Caribbean slab does not appear to extend north of the Sierra de Bahoruco. The lithospheric mantle underneath the central part of Hispaniola has apparently disappeared between these regions, although we could not unambiguously explain it by the poor resolution of our data in that region.

Nevertheless, our results reveal a strong lateral velocity variation providing an unusual seismic structure where the EPGFZ is located. This velocity variation is evidenced by strong reflections associated with low-velocity zones that could be the expression of areas of enhanced pore pressure in the presence of a strike-slip fault zone (e.g., the Alpine Fault Zone in Stern and McBride, 1998). In this anomalous domain, the velocity in the upper mantle is 7.0–7.5 km/s while the adjacent areas are 7.6–8.0 km/s. This anomaly could be related to the collision region between both tectonic plates. The Engdahl catalog of well-located earthquakes and the seismicity reported by the ISC (Engdahl et al., 1998) in the western region of Hispaniola (Fig. 6) show that both in the north and the south of our profile, the events are concentrated in the main fault zones. In the central part, however, there seems to be no seismicity. These regions correspond to the two main strike-slip fault zones between which the fold and thrust belts are located, indicating a state of compressive stresses.

In southern Hispaniola, the Enriquillo basin is filled by Pleistocene sedimentary rocks (Braga et al., 2012). Mann et al. (2002) proposed a thickness by of 5 km and a Moho depth at 26 km, in agreement with the values obtained in this study (5 km and 27 km, respectively). Also, these authors established that the beginning of the deformation in the region of the San Juan-Azúa ramp basin do not present thinning towards the north or the south in the sedimentary units. However, the current pattern of morphotectonic units in the central Hispaniola is associated with late Miocene and younger reverse and oblique-slip faulting, including the Enriquillo, San Juan-Azúa, and Cibao thrust-bound basins. In the southern Haiti portion of the EPGFZ, Douilly et al. (2016) suggest the existence of low-velocity structures delineated by sedimentary basins and fault structures similar to the sedimentary basin present in our study area. This pattern is in agreement with the southwest-verging reverse faulting movement; the EPGFZ northern flank is the steeper dipping flank of the San Juan-Azúa basin. Our model proposes that in the eastern coastal part of the study area, the crust thickness is 27 km increasing in the region of Bahía de Neiba up to 32 km and thinning across the Muertos Trough down to 17 km. In this region, our data show that the maximum depths are between 43–47 km and the average velocity is 8.1 km/s.

In the Caribbean Sea region, sub-crustal tectonic features of the Hispaniola-Caribbean plate collision zone reveal a thinner crust in the Haiti sub-basin (5–10 km) (Granja Bruña et al., 2014; Mann et al., 2002). This region is also affected by the indentation of the Beata Ridge close to the Haiti sub-basin, which produces variations in the seismic velocity, thickness of the crust layers (Núñez et al., 2016), and seismicity under the Sierra de Bahoruco (Rodríguez et al., 2018). Furthermore, the proposed model in this work is characterized by 4 km thick of highly reflecting layers in the Haiti sub-basin that would correspond to ponded sediments. The P-wave velocities obtained in this work and by Núñez et al. (2016) are in agreement with previous studies. Those studies suggested for this region of the Caribbean Sea crustal thickness and velocities similar to those of many other oceanic plateaus including the Caribbean plateau (Driscoll and Diebold, 1999; James, 2009; Edgar et al., 1971). The deep structure obtained in this study is characterized by crustal and upper mantle deformation related to the presence of the Muertos Trough and the Beata Ridge.

Table 2
Moho boundary depth accuracy from the comparison between various lower crust velocity (LCV) and modeled in km/s. The values represent the differences between Moho depth of P-wave velocity model proposed in this paper and Moho depths obtained from different LCVs along: i) Northern region: MB Mouchair Bank, PPCoast Puerto Plata Coast, PPC Puerto Plata Complex, CS Cordillera Septentrional, CV Cibao Valley; and ii) Southern region: SB Sierra de Bahoruco, SC Southern Coast, HsB Haiti sub-Basin.

| | Northern region | | | | | Southern region | | | |
|---|---|---|---|---|---|---|---|---|---|
| LCV (km/s) | MB, PPCoast 25 – 75 (km) | PPC, CS, CV 75-150 (km) | CC 150-200 (km) | Total Average (km) | | LCV (km/s) | SB, SC 300-350 (km) | HsB 350-450 (km) | Total Average (km) |
| 6.3 | −0.32 | −2.67 | −1.25 | −0.43 | | 6.4 | −0.3 | −0.38 | −0.35 |
| 6.7 | 0.52 | 4.2 | 0.28 | 1.99 | | 6.8 | 1.72 | 1.36 | 1.49 |
| 7.0 | 2.35 | 9.54 | 5.08 | 6.03 | | 7.0 | 8.1 | 3.6 | 4.4 |

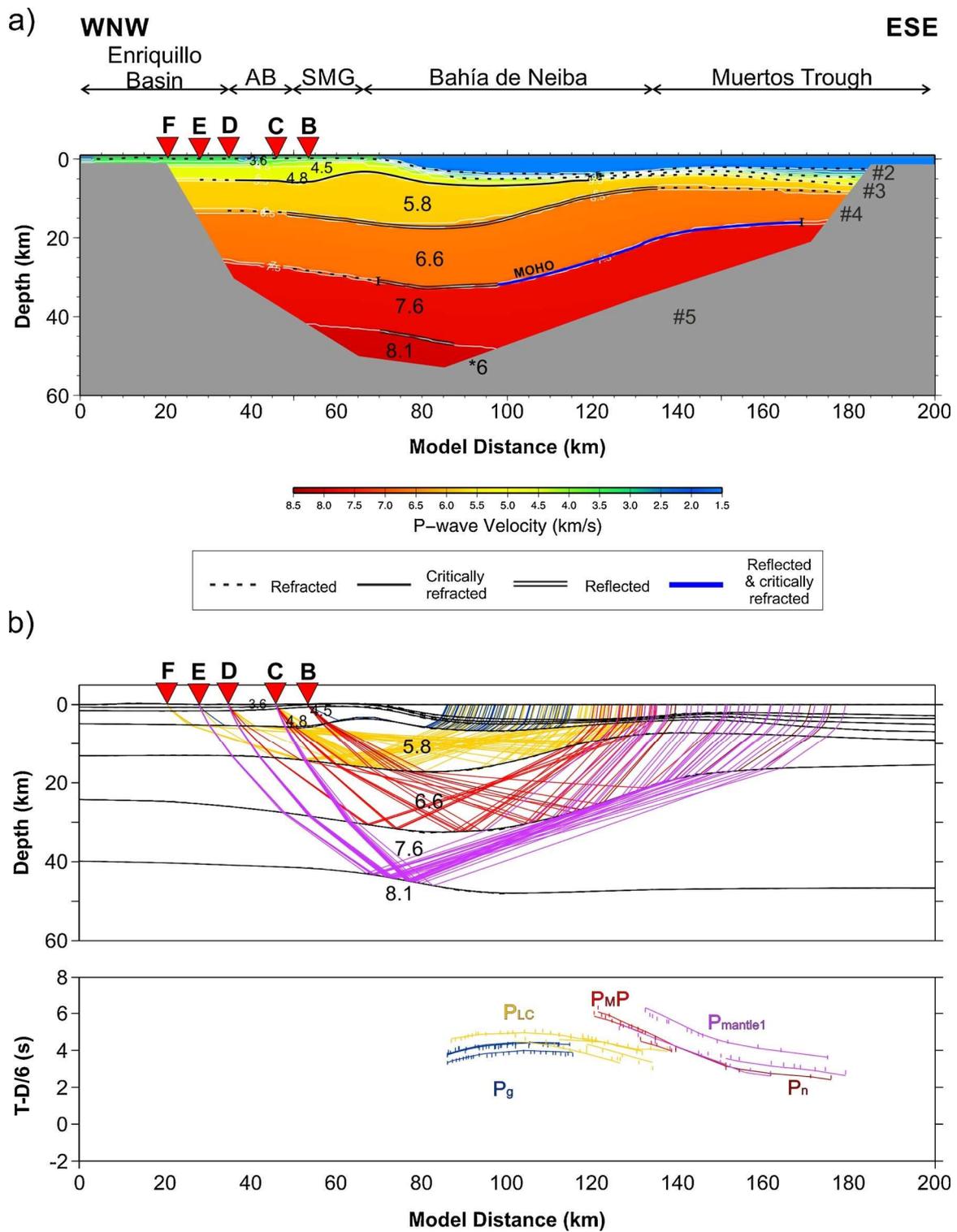

Fig. 5. (a) P-wave velocity model across the southwestern part of the Dominican Republic (GEOPRICO-DO Project. Seismic stations of Profile F are depicted by red inverted triangles. Horizontal and vertical axis show model position and depth below 0 in km, respectively. The colored area shows the domain where ray tracing limits velocities. The lines represent the boundary of the layers, while the positions where the rays are refracted, critically refracted, reflected and both simultaneously are marked by pointed, solid, double and blue thick lines, respectively, showing the well-defined areas. The thick grey line represents the reflections over the floating reflector. The uncertainty in depth of Moho discontinuity is represented by vertical bars. The region not crossed by rays is represented by a grey area, while the question marks show regions with poor control over the ray coverage. The layer numbers are identified by #N, where N is the layer number, and N* depicts floating reflectors. (b) Top: Ray tracing and velocity model with average velocity in the corresponding layer in km/s. Bottom: Comparison between observed (vertical bars) and calculated (lines) travel times. Its height is the uncertainty estimated for every phase. In all plots, distances refer to the velocity model origin. AB Azúa Basin, BN Bahía de Neiba, EB Enriquillo Basin, MT Muertos Trough, SMG Sierra de Martín García. (For interpretation of the references to colour in this figure legend, the reader is referred to the web version of this article.)

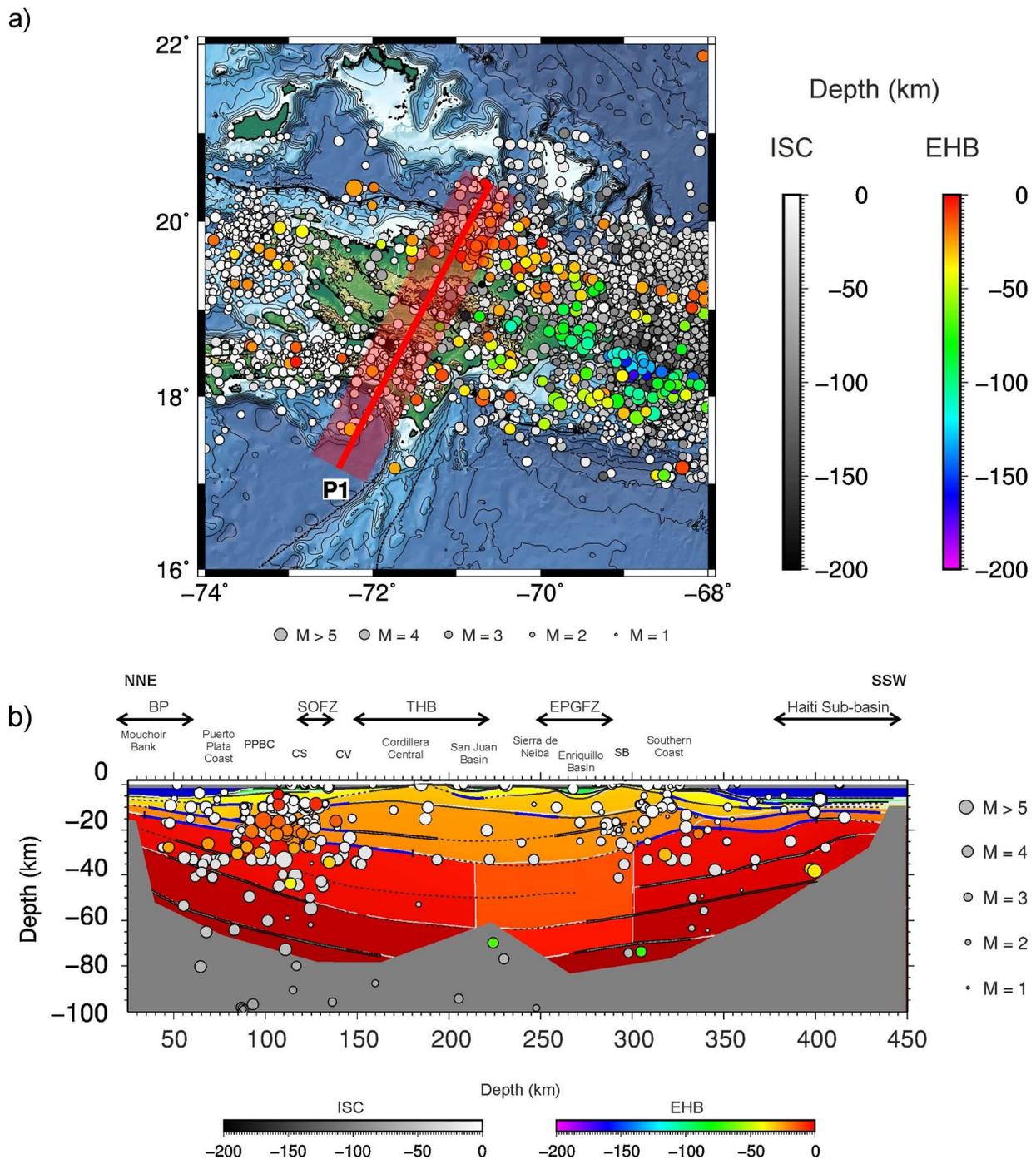

Fig. 6. a) Seismicity from 1950 to 2016 in grey scale (International Seismological Centre, 2011) and the Engdahl' catalog of well-located earthquakes from 1960 to 2008 (color scale) in the Hispaniola Island region (Engdahl et al., 1998). Profile P1 (red line) represents the projection of the seismicity along the Profile A seismic profile with a width of 30 km (red shadow band). b) Profile A P-wave velocity model with seismicity projected along profile P1. Abbreviations as in Fig. 4a. (For interpretation of the references to colour in this figure legend, the reader is referred to the web version of this article.)

According to the model proposed by Corbeau et al. (2017), we can establish correspondences and discrepancies between the results of both studies. First, these studies do not cover the exact same areas and tectonic units. Since, these authors used teleseismic data with Mw > 5 and located at approximate distances of 3000 to 10,000 km between the sources and the receivers, the dataset used are significantly different with locations clearly biased towards the south of their study area. The two studies distinguish three different regions: north, south and central domains (Fig. 7). Both in the northern and the southern regions, Corbeau et al (2017) obtain Moho depths that are in agreement with the values proposed in this work. However, significant differences were observed in the central domain, from the south of the Cordillera Central to the north of the Sierra de Bahoruco. Our results provide an important lateral velocity change reaching depths between 50 and 60 km and constrained by seismic data collected in situ in the study area. Corbeau et al. (2017), however, obtain a pronounced thickening of the crust (32–45 km) that would be formed by relatively similar materials and would not produce significant changes in the seismic velocity.

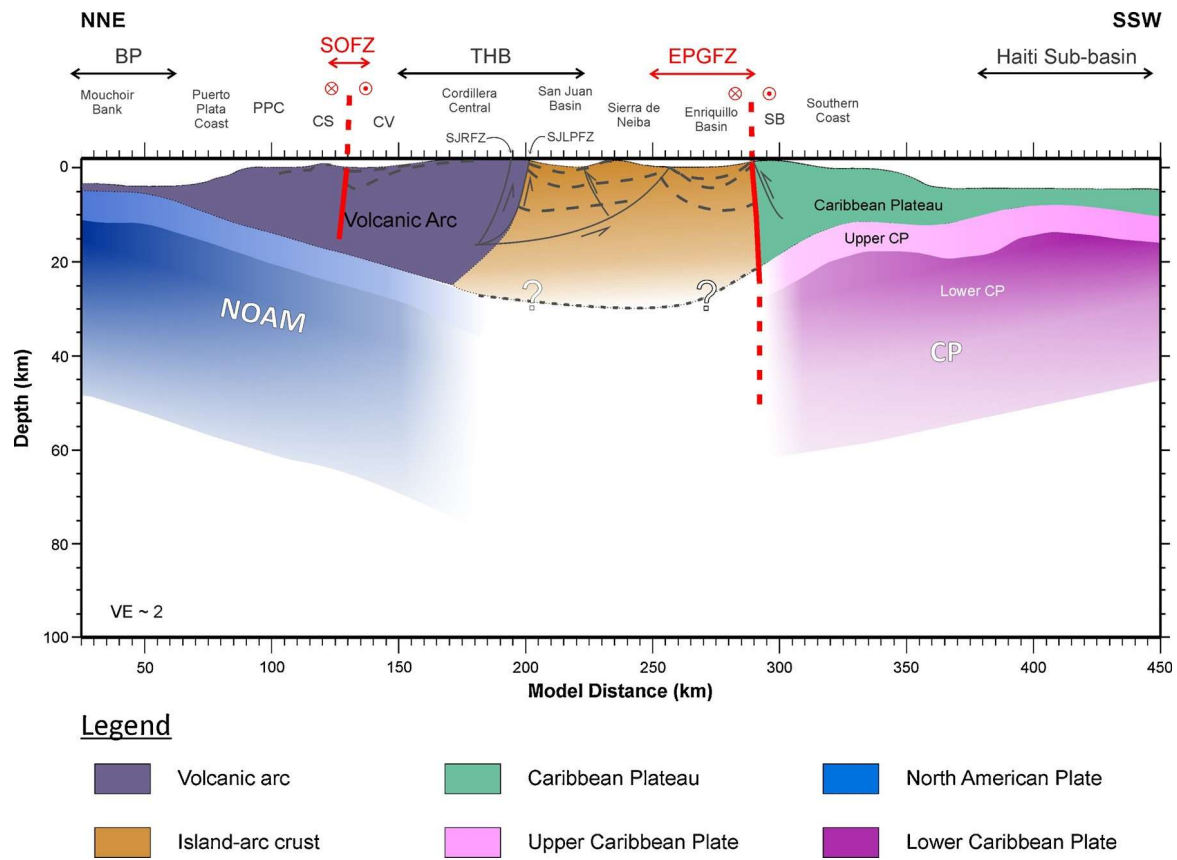

Fig. 7. Conceptual tectonic scheme summarizing the main results obtained from Profile A (see Fig. 2 for localization of the profile) including differentiation in regions proposed by Corbeau et al. (2017) and the tectonic faults (in grey) by de Zoeten and Mann (1999). Question marks denote hypothetical locations of NOAM, CP and Moho discontinuity depth proposed in this work. Abbreviations as in Fig. 4a. LIP Large Igneous Province, SJRFZ San José – Restauración Fault Zone, SJLPFZ San Juan – Los Pozos Fault Zone. (For interpretation of the references to colour in this figure legend, the reader is referred to the web version of this article.)

## 5. Conclusions

As a synthesis, our findings on the crustal structure in the west of the Dominican Republic may be enumerated as follow:

The cross-sectional lithospheric models for Profile A and F with NNE-SSW and WNW-ESE directions, respectively, consists of seismic layers grouped in upper, middle and lower crust and uppermost mantle:
i) The upper crust is characterized as the most inhomogeneous part of the model due to the succession of sedimentary basins and mountain ranges formed by upper and lower sediments. The southern part is characterized by lateral velocity variations related to the indentation of the Beata Ridge, which also has an observable impact within the middle crust, and disappearing in the vicinity of the Muertos Trough. ii) The middle crust corresponds to the upper transitional crust and is thinned in the areas of the Cordillera Central, the Sierra de Bahoruco and the Muertos Trough in the SE and slightly thickened in the Puerto Plata Complex, Enriquillo Plantain Garden Fault Zone and Caribbean coast of the studied area. iii) The lower crust is related to lower transitional crust in the onshore part. This layer is homogeneous in velocity from N to S of the model except in the area of the Caribbean Sea, where a lateral velocity variation due to the presence of Beata Ridge is observed.

The depth of the Moho discontinuity was determined and displayed a strong lateral variation beneath the Enriquillo Plantain Garden Fault Zone. In the Bahía de Neiba area, the crust is thinner due to the presence of the Muertos Trough.

In the NNE area, the Moho and upper mantle layers exhibit a dipping of about 11° becoming deeper because of the north-dipping slab towards the interior of the island while the southern slab is deformed by contact between the Caribbean plate and Hispaniola with an average dip angle 8°–9°.

The wide-angle seismic data used in this study do not reveal significant changes in the crust and upper mantle structure coinciding with the Septentrional Fault Zone in the area of the western flank of the Cordillera Septentrional.

## Declaration of Competing Interest

None.

## Acknowledgments and Data

The authors would especially like to thank Raphael de Plaen, Felipe de Jesús Escalona-Alcázar and Francisco J. Núñez-Cornú for their invaluable help. We are also very grateful with the Editor, Ling Chen, and the anonymous reviewers for their constructive comments and suggestions that improve this work considerably. This research was funded by the following projects: CTM2006-13666-C02-02 (CARIBE NORTE), CTM2008-02955-E/MAR, CGL2011-29474-C02-01 (TSUJAL) and RTI2018-094827-B-C21 (KUK AHPAN). The authors thank the BIO Hespérides Captain and crew, UTM-CSIC technicians, the CARIBE NORTE working group, and IRIS-PASSCAL Instrument Center for lending us 300 seismic stations (Experiment Code 0921). The authors would like to thank the General Department of Mining, University Seismological Institute (ISU), Autonomous University of Santo Domingo (UASD), and Dominican Navy who provided two coastguard ships 106-BELLATRIX and 109-ORION for their collaboration during the data acquisition. D. Núñez was financially supported by a doctoral fellowship from the Spanish Ministry of Education and Science (BES-2008-001997). All geophysical data collected in CARIBE NORTE and GEOPRICO-DO Projects, once finished, will be in a database at Research